\documentclass[ floats, showpacs,nofootinbib,floatfix,prl,onecolumn]{revtex4}%
\usepackage{eurosym}
\usepackage{amsfonts}
\usepackage{amsmath}
\usepackage{graphicx}
\usepackage{subfigure}
\usepackage{amssymb}%
{\catcode`\|=\active \gdef\Braket#1{\left<\mathcode`\|"8000\let|\bravert
{#1}\right>}}
\def\bravert{\egroup\,\vrule\,\bgroup}

\newcommand{\beq}{\begin{eqnarray}}
\newcommand{\eeq}{\end{eqnarray}}
\begin{document}
\title{Enhancement effects in exclusive $\pi\pi$ \ and $\rho\pi$ production in
$\gamma^{\ast}\gamma$ scattering}
\author{A. Courtoy}
\email{Aurore.Courtoy@uv.es}
\author{S. Noguera}
\email{Santiago.Noguera@uv.es}
\affiliation{Departamento de F\'{\i}sica Te\'orica and Instituto de F\'{\i}sica
Corpuscular, Universidad de Valencia-CSIC, E-46100 Burjassot (Valencia), Spain.}
\date{\today }

\begin{abstract}
The exclusive $\pi\pi$ \ and $\rho\pi$ production in hard $\gamma^{\ast}%
\gamma$ scattering in the forward kinematical region where the virtual photon
is highly off-shell is studied using the $\gamma\to\pi^-$ Transition Distribution Amplitudes
obtained in realistic models for the pion. For $\rho\pi$
production we confirm the previous estimates before QCD evolution. Nevertheless,
once evolution is taken into account this cross section grows one order of
magnitude. In the case of $\pi\pi$ production we have evaluated the cross
section including the pion pole contribution. We observe that this
contribution is responsible for an enhancement of two orders of magnitude with
respect to the cross section evaluated without the pion pole term.

\end{abstract}

\pacs{11.10.St, 12.38.Lg, 13.60.-r, 24.10.Jv}
\maketitle

\begin{figure}
[tb]
\begin{center}
\includegraphics[
trim=0.000000cm 0.000000cm 0.000000cm 0.078035cm,
height=5.0215cm,
width=5.3554cm
]%
{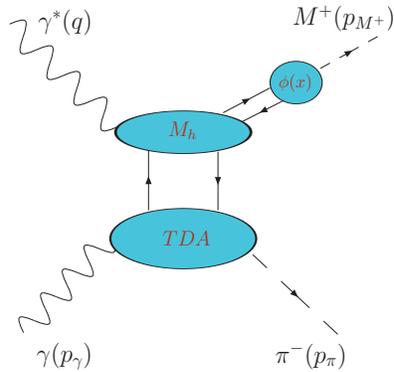}%
\caption{Factorization diagram defining the TDA for the process $\gamma^{\ast
}\gamma\rightarrow\pi M$ at small momentum transfer, $t=\left(  q-p_{M}%
\right)  ^{2},$ and large invariant mass, $s=\left(  p_{\gamma}+q\right)
^{2}$.}%
\label{facto}%
\end{center}
\end{figure}

Collisions of a real photon and a highly virtual photon are an useful tool for
studying fundamental aspects of QCD. Inside this class of processes, the
exclusive meson pair production in $\gamma^{\ast}\gamma$ scattering has been
analyzed in Ref.~\cite{Pire:2004ie} introducing of a new kind of distribution
amplitudes, called Transition Distribution Amplitude (TDA). At small momentum
transfer $t$ and in the kinematical regime where the photon is highly virtual,
a separation between the perturbative and the nonperturbative regimes is
assumed to be valid. Through the factorization theorems the amplitude for such
reactions can be written as a convolution of a hard part $M_{h}$, the meson
distribution amplitude $\phi_{M}$ and a soft part, the TDA, describing the
photon-pion transition, as is shown in Fig.~\ref{facto}. Lacking a complete
fundamental understanding of the color dynamics, we are compelled to use
models for predictions. Cross section estimates for the processes
\begin{equation}
\gamma_{L}^{\ast}\gamma\rightarrow\pi^{+}\pi^{-}\ \ ,\quad\gamma_{L}^{\ast}
\gamma\rightarrow\rho^{+}\pi^{-} \quad,
\label{processes}%
\end{equation}
have been proposed in Ref.~\cite{Lansberg:2006fv} using for the TDA a
$t$-independent double distributions, in a first approach, and, in a second,
the $t$-dependent results of Ref.~\cite{Tiburzi:2005nj}.%

The process $\gamma^{\ast}\gamma\rightarrow\pi\pi$ is particularly interesting
because different kinematical regimes lead to different mechanisms, which
implies a description of the process through either the pion Generalized 
Distribution Amplitudes or the pion-photon 
TDAs \cite{Anikin:2008bq}. Therefore, it is of interest to
deepen our understanding of the
description of this process. Recently the pion-photon TDAs have been
calculated in, respectively, the Spectral Quark Model (SQM)
\cite{Broniowski:2007fs}, the Nambu - Jona-Lasinio model with Pauli-Villars
regularization procedure (NJL) \cite{Courtoy:2007vy} and a nonlocal chiral
quark model \cite{Kotko:2008gy}. A comparison of the results obtained in these
three models is given in Ref.~\cite{Courtoy:2008ij} concluding that there is
clear agreement between the different studies of the pion-photon TDAs,
allowing us to analyze the result of a single model, e.g. the NJL model. Our
choice is based on the fact that the NJL model \cite{Nambu:1961tp} is the most
realistic model for the pion based on a local quantum field theory built with
quarks. It gives a right description of the low energy pion physics. It has
been applied to the study of pion parton distribution (PD)
\cite{Davidson:1994uv} and pion generalized parton distribution (GPD)
\cite{Theussl:2002xp}. In the chiral limit, its quark valence distribution is
as simple as $q(x)=\theta\left(  x\right)  ~\theta(1-x)$. Once evolution is
taken into account, good agreement is reached between the calculated PDF and
the experimental one \cite{Davidson:1994uv}. More elaborated studies of pion
PD have been done in the Instanton Liquid Model \cite{Anikin:2000rq} and
lattice calculation based models \cite{Noguera:2005cc} using nonlocal
Lagrangians \cite{Noguera:2005ej}, which confirm that the result obtained in
the NJL model for the PD is a good approximation. The QCD evolution of the
pion \ GPD calculated in the NJL model has been also studied in
\cite{Broniowski:2007si}.

Before defining the TDAs, we introduce the light-cone coordinates
$v^{\pm}=\left(  v^{0}\pm v^{3}\right)  /\sqrt{2}$ and the\textbf{ }transverse
components $v^{\bot}=\left(  v^{1},v^{2}\right)  $ for any four-vector
$v^{\mu}$. We define $P=\left(  p_{\pi}+p_{\gamma}\right)  /2$\ and we
introduce the light-front vectors $\bar{p}^{\mu}=P^{+}\left(  1,0,0,1\right)
/\sqrt{2}$ and $n^{\mu}=\left(  1,0,0,-1\right)  /\left(  \sqrt{2}%
P^{+}\right)  .$ The momentum transfer is $\Delta=p_{\pi}-p_{\gamma},$
therefore $P^{2}=m_{\pi}^{2}/2-t/4$ and $t=\Delta^{2}$. The skewness variable
describes the loss of plus momentum of the incident photon, i.e. $\xi=\left(
p_{\gamma}-p_{\pi}\right)  ^{+}/2P^{+},$ and its value ranges between
$-1<\xi<-t/\left(  2m_{\pi}^{2}-t\right)  $. With these conventions, the
vector and axial TDA are defined by
\begin{align}
\int\frac{dz^{-}}{2\pi}e^{ixP^{+}z^{-}}\left.  \left\langle \pi^{\pm}(p_{\pi
})\right\vert \bar{q}\left(  -\frac{z}{2}\right)  \gamma^{+}\hspace*{0pt}%
\tau^{\pm}\ q\left(  \frac{z}{2}\right)  \left\vert \gamma(p_{\gamma
}\varepsilon)\right\rangle \right\vert _{z^{+}=z^{\bot}=0}  &  =\frac{1}%
{P^{+}}\ i\,e\,\varepsilon_{\nu}\,\epsilon^{+\nu\rho\sigma}\,P_{\rho}%
\,(p_{\pi}-p_{\gamma})_{\sigma}\,\frac{V^{\gamma\rightarrow\pi^{\pm}}%
(x,\xi,t)}{\sqrt{2}f_{\pi}}\quad,\nonumber\\
& \label{vectortda}\\
\int\frac{dz^{-}}{2\pi}e^{ixP^{+}z^{-}}\left.  \left\langle \pi^{\pm}\left(
p_{\pi}\right)  \right\vert \bar{q}\left(  -\frac{z}{2}\right)  \gamma
^{+}\hspace*{-0.05cm}\gamma_{5}\ \tau^{\pm}\ q\left(  \frac{z}{2}\right)
\left\vert \gamma\left(  p_{\gamma}\varepsilon\right)  \right\rangle
\right\vert _{z^{+}=z^{\bot}=0}  &  =\pm\ \frac{1}{P^{+}}\left[  -\ e\,\left(
\vec{\varepsilon}^{\bot}\cdot(\vec{p}_{\pi}^{\bot}-\vec{p}_{\gamma}^{\bot
})\right)  \frac{A^{\gamma\rightarrow\pi^{\pm}}(x,\xi,t)}{\sqrt{2}f_{\pi}%
}\right. \nonumber\\
&  \left.  +\ e\,\left(  \varepsilon\cdot(p_{\pi}-p_{\gamma})\right)
\frac{2\sqrt{2}f_{\pi}}{m_{\pi}^{2}-t}~\epsilon(\xi)~\phi_{\pi}\left(
\frac{x+\xi}{2\xi}\right)  \right]  \quad. \label{axialtda}%
\end{align}
where the pion decay constant is $f_{\pi}=92.4.$ MeV, $\epsilon\left(
\xi\right)  $ is equal to $1$ for $\xi>0$ and to $-1$ for $\xi<0$ and
$\phi_{\pi}(x)$ is the pion DA. Here we have modified the definition given in
Refs.~\cite{Pire:2004ie,Lansberg:2006fv} in order to introduce the pion pole
contribution \cite{Tiburzi:2005nj, Courtoy:2007vy} in Eq.~(\ref{axialtda}).
 This equation deserves some comments. The pion pole term in (\ref{axialtda})
 describes a point-like pion propagator multiplied by the distribution
amplitude (DA) of an on-shell pion. It contributes to the axial current
through a different momentum structure and must be subtracted in order to
obtain de axial TDA. We emphasize that it is a model
independent definition, because we have define the numerator of the pion pole
term as the residue at the pole $t=m_{\pi}^{2}$. With this definition, all the
structure dependence related to the outgoing $\pi^{\pm}$ is included in
$A\left(  x,\xi,t\right)$. Moreover, the pion pole contribution can be
estimated in a phenomenological way, as we will see later on.\ With these
definitions we recover the sum rules
\begin{equation}
\int_{-1}^{1}dx\ D\left(  x,\xi,t\right)  =\frac{\sqrt{2}\ f_{\pi}}{m_{\pi}%
}\ F_{D}\left(  t\right)  \ \ \ ,\ \ \ \ \ \ \ \ \ \ \ \ \ \ \ D=V,A\quad,
\label{sumrules}
\end{equation}
with the standard definitions for the form factors $F_{V,A}$ appearing in the
$\pi^{\pm}\rightarrow\ell^{\pm}\nu\gamma$ decay \cite{Amsler:2008zz}. Notice
that the on-shell pion DA obeys the normalization condition $\int_{0}%
^{1}dx\,\phi_{\pi}(x)=1$.

The $\gamma^{\ast}\gamma\rightarrow M^{+}\pi^{-}$ process$,$ with $M^{+}%
=\rho_{L}^{+}$ or $\pi^{+},$ is a subprocess of the $e\left(  p_{e}\right)
+\gamma\left(  p_{\gamma}\right)  \rightarrow e\left(  p_{e}^{\prime}\right)
+M^{+}\left(  p_{M}\right)  +\pi^{-}\left(  p_{\pi}\right)  $ process. We
follow all the definitions of the kinematics given in the section III.A and
Fig. 3 of Ref.~\cite{Lansberg:2006fv}, with the exception that our $n^{\mu}$
vector is twice the $n^{\mu}$ vector used in \cite{Lansberg:2006fv} (i.e. $n.p=1$
with our definitions). In particular, for massless pions, we have
\begin{equation}
Q^{2}=-q^{2}=-\left(  p_{e}-p_{e}^{\prime}\right)  ^{2}%
\ \ ,\ \ \ \ \ \ \ \ \ s_{e\gamma}=\left(  p_{e}+p_{\gamma}\right)  ^{2}\quad,
\end{equation}
\begin{align}
p_{\gamma}=(1+\xi)\bar{p}\quad ,&\quad  p_{\pi}=(1-\xi)\bar{p}+\frac{\vec{\Delta
}^{\perp2}}{2\left(  1-\xi\right)  }n+\vec{\Delta}^{\perp},\nonumber\\
q  &  =-2\xi\bar{p}+\frac{Q^{2}}{4\xi}n\quad,
\end{align}
where $\Delta_{T}=(0,\vec{\Delta}^{\perp},0)$ and therefore $\Delta_{T}%
^{2}=-\vec{\Delta}^{\perp2}$. Notice that $\vec{\Delta}^{\perp2}%
=(-t)(1-\xi)/(1+\xi)$, with $t<0$. The longitudinal polarization of the
incoming virtual photon is 
$\varepsilon_{L}=\left(2\xi\bar{p}/Q +Q n/(4\xi)\right)$.
The real photon polarization is defined by the condition
 $\varepsilon^{-}=0$ together with the gauge
condition $\varepsilon^{+}=0$.

The differential cross section is given by\footnote{A
factor $1/4$ is missing in Eq.~(23) of Ref.~\cite{Lansberg:2006fv}. This
typo does not affect to the numerical results reported in that Reference \cite{jp}.} \cite{Lansberg:2006fv}
\begin{equation}
\frac{d\sigma^{e\gamma\rightarrow eM^{+}\pi^{-}}}{dQ^{2}dtd\xi}=\frac
{64\,\pi^{2}\ \alpha_{em}^{3}}{9\,s_{e\gamma}\ Q^{8}}\,\frac{\left(
1-\xi\right)  }{(1+\xi)^{4}}\,\left(  2\xi s_{e\gamma}-Q^{2}\left(
1+\xi\right)  \right)  \left(  -t\right)  \,\left\vert \mathcal{I}%
_{M}\right\vert ^{2}\quad, \label{CroosSecc}%
\end{equation}
with
\begin{align}
\mathcal{I}_{\rho}  &  =\frac{\alpha_{s}}{6}\,\int_{-1}^{1}dx\,\int_{0}%
^{1}dz\,\frac{f_{\rho}}{f_{\pi}}\,\phi_{\rho}(z)\,\frac{1}{z\left(
1-z\right)  }\left(  \frac{Q_{u}}{x-\xi+i\epsilon}+\frac{Q_{d}}{x+\xi
-i\epsilon}\right)  \,V^{\gamma\rightarrow\pi^{-}}(x,\xi,t)\quad,\label{rho}\\
\mathcal{I}_{\pi}  &  =\frac{\alpha_{s}}{6}\,\int_{-1}^{1}dx\,\int_{0}%
^{1}dz\,\phi_{\pi}(z)\,\frac{1}{z\left(  1-z\right)  }\left(  \frac{Q_{u}%
}{x-\xi+i\epsilon}+\frac{Q_{d}}{x+\xi-i\epsilon}\right)  \,\left[
A^{\gamma\rightarrow\pi^{-}}(x,\xi,t)-\frac{4\,f_{\pi}^{2}}{t-m_{\pi}^{2}%
}\,\epsilon\left(  \xi\right)  \,\phi_{\pi}\left(  \frac{x+\xi}{2\xi}\right)
\right] \,,\nonumber\\
& \label{pi}
\end{align}
where $z$ is the light-cone momentum fraction carried by the quark entering
the meson $M^{+}$ and $f_{\rho}=0.216%
\operatorname{GeV}%
$  and $Q_{q}$ is the electric charge of the quark $q$. The last term on the r.h.s. of Eq.~(\ref{pi}) is the pion pole contribution
to the amplitude coming from the second term of Eq.~(\ref{axialtda}).

We proceed now to evaluate these integrals. The meson DA is chosen to be the usual asymptotic normalized meson DA, i.e.
$\phi_{M}(z)=6z(1-z)$, which cancels the $z$-dependence of the hard amplitude.
For the nonperturbative part of the process we use the TDAs evaluated in the
NJL model. This approach is based on the determination of the pion as a bound
state through the Bethe-Salpeter equation. This guarantees that all
invariances of the problem are preserved. As a consequence, the obtained TDAs
explicitly verify the sum rules, the polynomiality condition, the isospin
relations and have the correct support in $x$ \cite{Courtoy:2007vy}.

In Ref.~\cite{Courtoy:2007vy} the $\pi^{+}\rightarrow\gamma$ TDAs were
calculated, which are connected to the $\gamma\rightarrow\pi^{-}$ TDAs through
$CPT$ symmetry \cite{Courtoy:2008ij}
\begin{equation}
D^{\gamma\rightarrow\pi^{-}}\left(  x,\xi,t\right)  =D^{\pi^{+}\rightarrow
\gamma}\left(  -x,-\xi,t\right)  \quad, \ \ \ \ \ \ \ \ \ \ \ \ \ \ D=V,A\quad.
\end{equation}

\begin{figure}
[ptb]
\begin{center}
\includegraphics[
trim=0.697746cm 0.698118cm 0.699478cm 0.699060cm,
height=8.1385cm,
width=16.0705cm
]%
{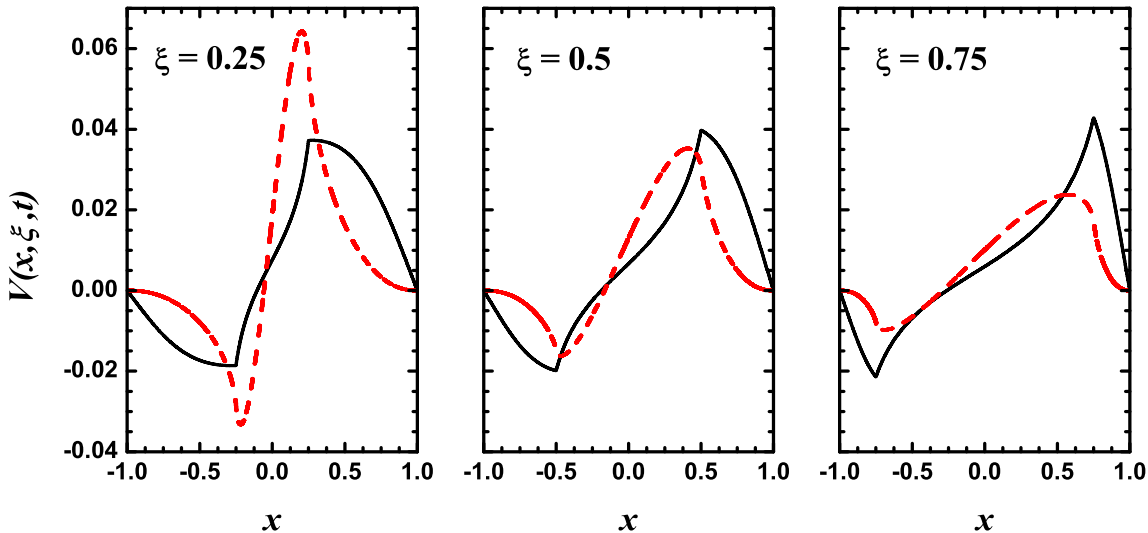}%
\caption{The functions $V^{\gamma\rightarrow\pi^{-}}(x,\xi,t)$ and for
$\xi=0.25,$ $0.5,$ $0.75$ and for $t=-0.5$ GeV$^{2}$. \ In each figure, the
solid line corresponds to the NJL model prediction and the dashed line to its
LO evolution.}%
\label{vTDA}%
\end{center}
\end{figure}

\begin{figure}
[ptbtb]
\begin{center}
\includegraphics[
trim=0.000000cm 0.000000cm 0.000000cm 0.860817cm,
height=5.6805cm,
width=8.96cm
]%
{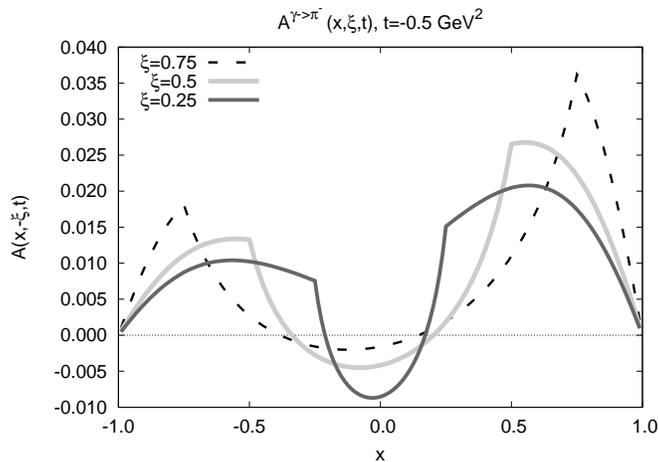}%
\caption{The functions $A^{\gamma\rightarrow\pi^{-}}(x,\xi,t)$ for different
values of the skewness variable $\xi$ and for $t=-0.5$ GeV$^{2}$. }%
\label{aTDA}%
\end{center}
\end{figure}

In Figs. \ref{vTDA} and \ref{aTDA} are shown the vector and axial $\gamma\to\pi^-$ TDAs for $t=-0.5$ for
different $\xi$ values. From Eq.~(\ref{CroosSecc}) it can be
observed  that $\xi\geq Q^{2}/\left( 2 s_{e\gamma}-Q^{2}\right)$. In other words, there is a (positive) lower limit on the value of $\xi$.
 It is indeed a particularly interesting  restriction because  the value of $\xi$ defines  the shape of the TDAs. 
In particular, the shape of the axial TDA radically changes according to the sign of the skewness variable;
 $A(x,\xi,t)$ has, at $x=\pm\xi$,
 its maximum values for $\xi>0$ (see Fig.~\ref{aTDA}) while it has its minimum values for $\xi<0$ \cite{Courtoy:2008ij}.
However the vector TDA has its maximum and minimum values at $x=\pm\xi$ (see Fig.~\ref{vTDA}) 
independently of the sign of the skewness variable. 
 On the other hand, the magnitude of the distributions is controlled by the $t$-dependence. This can be
easily understood because the TDAs, that  must satisfy the sum rules Eq.~(\ref{sumrules}),  are 
expected to decrease at least as $t^{-1}.$

\bigskip%

\begin{figure}
[ptb]
\begin{center}
\includegraphics[
trim=0.553911cm 0.555731cm 0.557113cm 0.556600cm,
height=6.8886cm,
width=13.4807cm
]%
{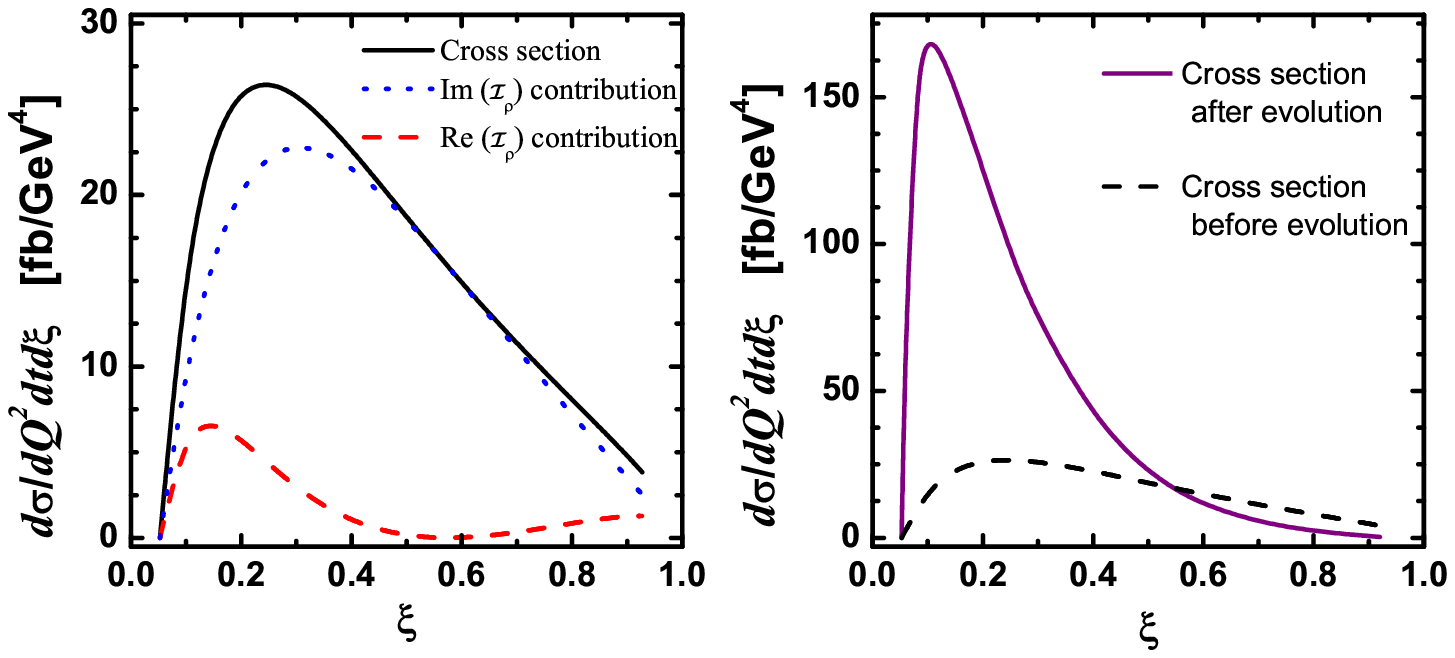}%
\caption{$e\gamma\rightarrow e^{\prime}\rho_{L}^{+}\pi^{-}$ differential cross
section plotted as a function of $\xi$ for $Q^{2}=4\operatorname{GeV}^{2},$
$s_{e\gamma}=40\operatorname{GeV}^{2},$ $t=-0.5\operatorname{GeV}^{2}.$ In the
first layer, the dotted\ (dashed) line is the contribution to the cross
section coming from the imaginary (real) part of the integral given in Eq.
(\ref{rho}). In the second layer\ we give the cross sections before and after
evolution.}%
\label{pi-rho-4}%
\end{center}
\end{figure}

In order to  numerically estimate the cross sections, we need to fix the strong coupling constant. 
In Ref.~\cite{Braun:2005be} is mentioned that a large value of $\alpha_{s}$
($\alpha_{s}\simeq1$) is indicated in the case of an asymptotic DA. Using
this value for $\alpha_{s}$\ we have evaluated the cross section for $\rho$
production. In Fig.~\ref{pi-rho-4} we plot this cross section as a function of
$\xi$. As we observe, the cross section is largely dominated by the imaginary
part of the integral of Eq.~(\ref{rho}). The $t$-dependence of the cross
section comes from both the overall $\left(  -t\right)  $ factor present in
Eq.~(\ref{CroosSecc}) and the $t$ dependence of the Vector TDA.
Therefore, we expect a decreasing as $t^{-1}$ for large $t$ values. Comparing
with the previous results in Ref.~\cite{Lansberg:2006fv}, we
observe that our predictions are higher by a factor 2 or 3.%

\begin{figure}
[ptb]
\begin{center}
\includegraphics[
trim=0.624743cm 0.625582cm 0.626528cm 0.626517cm,
height=6.5481cm,
width=13.3489cm
]%
{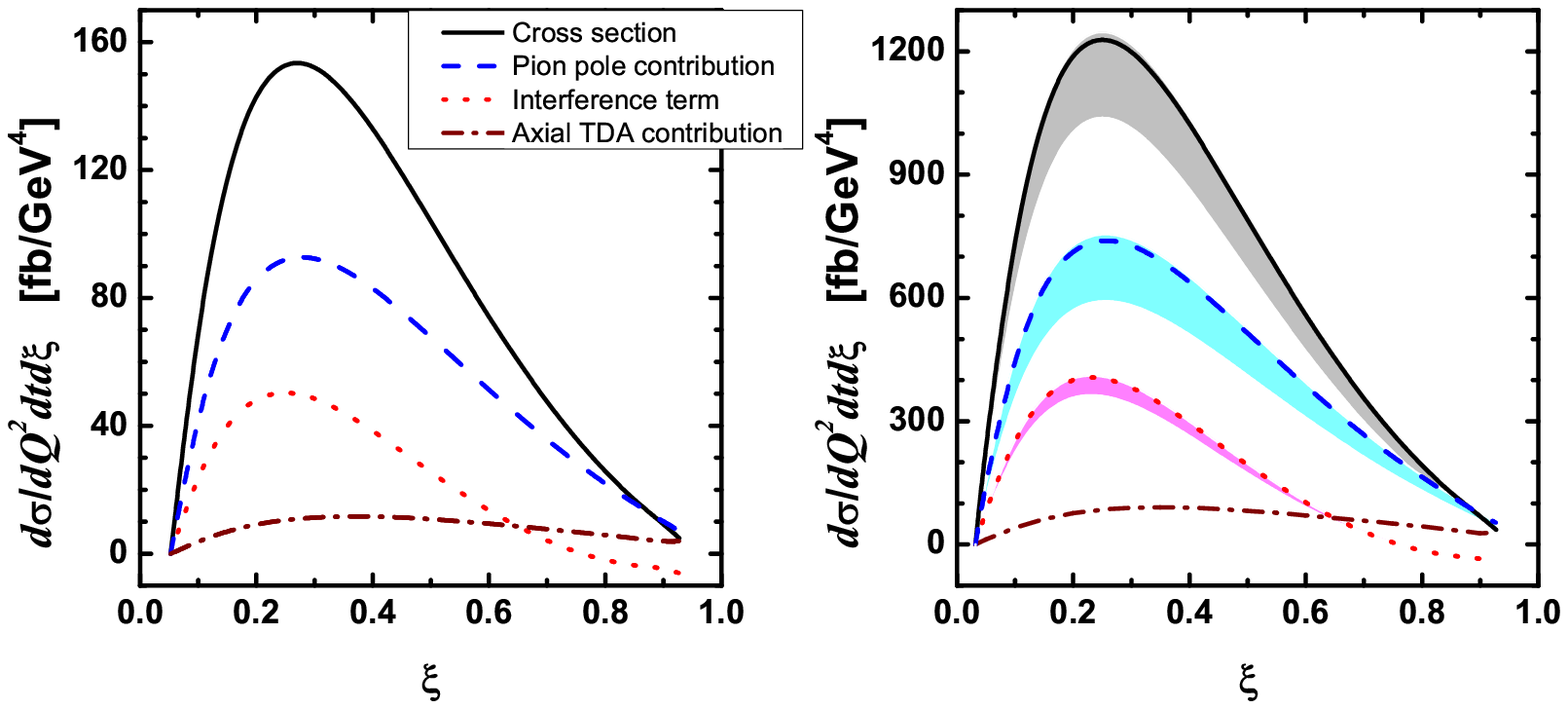}%
\caption{$e\gamma\rightarrow e^{\prime}\pi^{+}\pi^{-}$ differential cross
section plotted as a function of $\xi$ for $s_{e\gamma}=40\operatorname{GeV}%
^{2},$ $t=-0.5\operatorname{GeV}^{2},$ $Q^{2}=4\operatorname{GeV}^{2}$ (left)
and $Q^{2}=2.45\operatorname{GeV}^{2}$ (right)$.$ The
dashed\ (dashed-dotted)[dotted] line is the contribution to the cross section
coming from the pion fom factor (axial TDA) [interference term between the
pion FF and A]. The pion pole contribution is calculated using the
Brodsky-Lepage pion FF. The filled areas in the right layer correspond to the
same contributions but with the experimental value for the pion FF $F_{\pi
}=0.167\pm0.010$ \cite{Horn:2006tm}.}
\label{pi-pi-4}
\end{center}
\end{figure}

The $\pi$ production is described through Eq.~(\ref{pi}). Here we have a pion
pole term which becomes proportional to the electromagnetic pion form factor
(FF)
\begin{equation}
\mathcal{I}_{\pi}=\alpha_{s}\int_{-1}^{1}dx\,\left(  \frac{Q_{u}}
{x-\xi+i\epsilon}+\frac{Q_{d}}{x+\xi-i\epsilon}\right)  \,A^{\gamma
\rightarrow\pi^{-}}(x,\xi,t)-\frac{3}{4\pi}\frac{Q^{2}\,F_{\pi}\left(
Q^{2}\right)  }{t-m_{\pi}^{2}}
\quad.
\label{pi-FF}
\end{equation}
If we use the asymptotic form of the pion DA, i.e. $\phi_{\pi}\left(  z\right)  $
with $z=\left(  x+\xi\right)  /2\xi,$ in Eq.~(\ref{pi}) we obtain the
Brodsky-Lepage pion FF \cite{Lepage:1979zb}
\begin{equation}
Q^{2}\,F_{\pi}\left(  Q^{2}\right)  =16\pi\ \alpha_{s}\ f_{\pi}^{\,2}\quad.
\end{equation}
The cross section for pion production at $Q^{2}=4$ GeV$^2$ as a function of $\xi$ is given in
Fig.~\ref{pi-pi-4}(left). This cross section is dominated by the pion pole
contribution which is determined by the Brodsky-Lepage pion FF. Alternatively,
if the pion FF is experimentally known, we can infer phenomenologically this
contribution. And hence the axial TDA could be extracted from the interference term.
From Ref.~\cite{Horn:2006tm} we know that the pion FF at
$Q^{2}=2.45%
\operatorname{GeV}%
^{2}$ is $0.167\pm0.010$. In Fig.~\ref{pi-pi-4}(right) we depict, for each contribution, the prediction
 using the interval defined by the experimental value of the FF (filled areas), including also the  theoretical
prediction using the Brodsky-Lepage pion FF (lines).

The $t$-dependence of the cross section for pion production includes a strong
dependence on $t$ coming from the pion pole. Neglecting the pion mass in
(\ref{pi}), we observe that the pion pole contribution to $\mathcal{I}_{\pi}$
is proportional to $t^{-1}.$ Therefore, the cross section grows as $t^{-1}$
for small $t$ values. For large $t$ values we expect, as in the $\rho$ production
case, a decreasing as $t^{-1}.$%

\begin{figure}
[ptb]
\begin{center}
\includegraphics[
height=6.2516cm,
width=8.0309cm
]%
{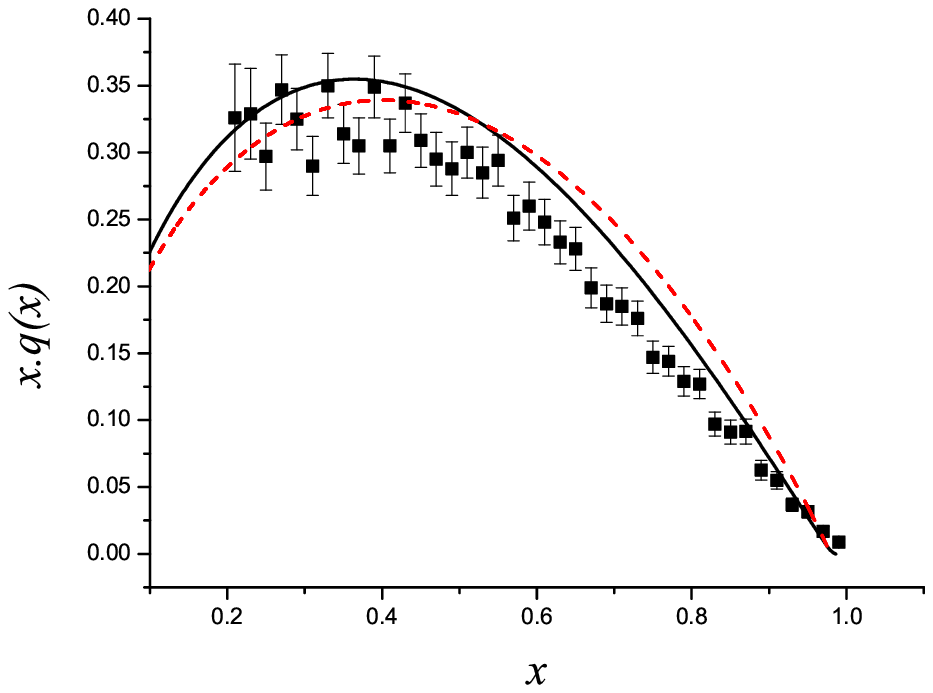}%
\caption{Pion parton distribution. The solid line corresponds to the LO
evolution of the NJL prediction and the dashed line to the NLO. Experimental
data are from \cite{Conway:1989fs}.}%
\label{PionPD}%
\end{center}
\end{figure}

We have also studied the effect of the QCD evolution on our estimates, using for this purpose the code of Freund
 and McDermott \cite{FreundMcDermott}.
One needs to fix the value of $Q_0$ for which  the quark distributions obtained in the NJL model are considered to be a good 
 approximation of the QCD quark distributions. Knowing that the
momentum fraction of each valence quark at $Q=2$ GeV is $0.23$ \cite{Sutton:1991ay}, we  fix the initial point of the evolution
in such a way that the evolution of the second moment of
the pion parton distribution reproduces this result. This condition is fulfilled
at a rather low value, i.e. $Q_{0}=0.29$ GeV, when the LO evolution is used. Going to the NLO  changes this value to
$Q_{0}=0.43$ GeV. However, in the latter case, the resulting PD is basically unaffected by the change in $Q$.
 The effect of the NLO evolution is compensated in the LO evolution
going to a lower value of $Q_{0}$,
 a result that has already been noticed
in  proton parton distributions \cite{Traini:1997jz}. In order to illustrate the latter statement,
 we have depicted both the evolved pion PD for the LO and
the NLO in Fig.~\ref{PionPD}.
Turning our attention to the vector TDA evolved at LO  (Fig. \ref{vTDA}), we observe that
the value of $V\left(  x,\xi,t\right)  $ at $x=\pm\xi$ grows for small values
of $\xi$ and decreases for large $\xi$ values in comparison with the TDA at the scale of the model.
This implies that the cross
section for $\rho$ production, which is largely dominated by the imaginary part,
will grow appreciably in the small $\xi$ region. In Fig.~\ref{pi-rho-4} we
compare the cross section after evolution, calculated only through
contribution of the imaginary part of $\mathcal{I}_{\rho},$ with the one
obtained before evolution. We observe that this cross section is multiplied by
a factor about 5 in the $\xi\sim0.2$ region.\footnote{The effect of QCD evolution, if calculated also through the contribution of the real part of  $\mathcal{I}_{\rho},$ could lead to a change in the cross section of about 15$\%$, what is within the model's uncertainties.} In the case of the axial TDA, the
cross section is dominated by the pion FF contribution, therefore the effect
of the evolution is expected to be small.

In many papers present in the literature, the r.h.s. of Eq.~(\ref{axialtda})
 contains only the $A$ term. In a general case the $A$ term
and the pion pole terms have different tensor structures, but we can fix the
gauge convention in such a way that these two structures coincide, as we have
done in this paper. In that case we can change our definition of the axial
TDA
\begin{equation}
\int\frac{dz^{-}}{2\pi}e^{ixP^{+}z^{-}}\left.  \left\langle \pi^{\pm}\left(
p_{\pi}\right)  \right\vert \bar{q}\left(  -\frac{z}{2}\right)  \gamma
^{+}\hspace*{-0.05cm}\gamma_{5}\ \tau^{\pm}\ q\left(  \frac{z}{2}\right)
\left\vert \gamma\left(  p_{\gamma}\varepsilon\right)  \right\rangle
\right\vert _{z^{+}=z^{\bot}=0}=\pm\ \frac{1}{P^{+}}\left[  -\ e\,\left(
\vec{\varepsilon}^{\bot}\cdot(\vec{p}_{\pi}^{\bot}-\vec{p}_{\gamma}^{\bot
})\right)  \frac{\bar{A}^{\gamma\rightarrow\pi^{\pm}}(x,\xi,t)}{\sqrt{2}%
f_{\pi}}\right]\,,
\end{equation}
with%
\begin{equation}
\bar{A}^{\gamma\rightarrow\pi^{\pm}}(x,\xi,t)=A^{\gamma\rightarrow\pi^{\pm}%
}(x,\xi,t)+\frac{4f_{\pi}^{2}}{m_{\pi}^{2}-t}~\epsilon(\xi)~\phi_{\pi}\left(
\frac{x+\xi}{2\xi}\right)  \ \ \ \ . \label{Abarra-A}%
\end{equation}
The latter expression shows that the pion pole contribution to the axial TDA
is closely related to the $D$-term of the Generalized Parton Distributions
\cite{Polyakov:1999gs}. Nevertheless, it must be realized that in this case it
gives an explicit contribution to the sum rule
\begin{equation}
\int_{-1}^{1}dx\ \bar{A}^{\gamma\rightarrow\pi^{\pm}}\left(  x,\xi,t\right)
=\frac{\sqrt{2}\ f_{\pi}}{m_{\pi}}\ F_{D}\left(  t\right)  +\frac{4f_{\pi}
^{2}}{m_{\pi}^{2}-t}\ \left(  p_{\gamma}-p_{\pi}\right)  .n\ \ \ \ .
\end{equation}

\bigskip

In this paper we have looked at the expected cross section for $\pi$-$\pi$
and $\pi$-$\rho$ production in exclusive $\gamma^{\ast}\gamma$ scattering in
the forward kinematical region using realistic models for the description of
the pion. First
we confirm the
previous estimates for $\rho$ production,
even if our results for the cross section are a factor 2 larger than the one
obtained in Ref.~\cite{Lansberg:2006fv}.
Second, in comparison with this previous evaluation of the cross
sections, we have improved in considering the effect of evolution on the vector current.
 In doing so,
 an additional factor 5 appears  in the small $\xi$ region,
leading to a cross section for $\rho$ production of one order of magnitude larger than the previous
calculation. We have also improved in including the pion pole term in the tensor decomposition of  the axial current.
 Then an even larger
enhancement factor, of about 60 in this case, is found in the cross section for pion production.
 The
interference term becomes a factor 15 larger than the pure TDA contribution,
making the axial TDA more accessible experimentally. The interest on TDAs is
actually extended to other transitions such as $\gamma^{\ast}N\rightarrow
N^{\prime}\pi,$ $\gamma^{\ast}N\rightarrow N^{\prime}\gamma,$ $N\bar
{N}\rightarrow\gamma^{\ast}\gamma,$ which are related to $N$-$\pi$ and
$N$-$\gamma$ transition distribution amplitudes \cite{Lansberg:2006uh,
Lansberg:2007ec, Pincetti:2008fh}.

\begin{acknowledgments}
We are thankful to J.P. Lansberg for useful discussions. This work has been
supported by the Sixth Framework Program of the European Commission under the
Contract No. 506078 (I3 Hadron Physics); by the MEC (Spain) under the Contract
FPA 2007-65748-C02-01 and the grant AP2005-5331 and by EU FEDER.
\end{acknowledgments}

\end{document}